\documentclass[aps,preprint,a4paper,12pt,nofootinbib]{revtex4-1}
\usepackage{lineno,hyperref}
\usepackage[utf8]{inputenc}
\modulolinenumbers[5]
\usepackage{graphicx}

\def\be{\begin{eqnarray}}
\def\ee{\end{eqnarray}}
\usepackage{amssymb}
\usepackage{rsfso}
\usepackage{eufrak}
\usepackage{yfonts}
\usepackage{xcolor}
\usepackage{dcolumn}

\begin{document}

\title{Analytical correspondence between shadow radius and black hole quasinormal frequencies}
\author{B. Cuadros-Melgar}
\email{bertha@usp.br}
\affiliation{Escola de Engenharia de Lorena, Universidade de S\~ao
   Paulo, Estrada Municipal do Campinho S/N, CEP 12602-810, Lorena, SP, Brazil}
\author{R. D. B. Fontana}
\email{rodrigo.fontana@uffs.edu.br}
\affiliation{Universidade Federal da Fronteira Sul, Campus Chapec\'o, CEP 89802-112, SC, Brazil}
\author{Jeferson de Oliveira}
\email{jeferson@gravitacao.org}
\affiliation{Instituto de F\'i­sica, Universidade Federal de Mato Grosso, CEP 78060-900, Cuiab\'a, MT, Brazil}




\begin{abstract}
We consider the equivalence of quasinormal modes and geodesic
quantities recently brought back due to the black hole shadow observation by
Event Horizon Telescope. Using WKB method we found an analytical
relation between the real part of quasinormal frequencies at the
eikonal limit and black hole shadow radius. We verify this
correspondence with two black hole families in $4$ and $D$ dimensions,
respectively.  
\end{abstract}


\maketitle


\section{Introduction}

Quasinormal modes exist as asymptotic solutions of propagating
fields (perturbations) around compact objects described by general
relativity or curvature based theories of gravitation. They are
characterized by a pair of numbers, a frequency of oscillation of such
a system together with its damping, $\omega = \omega_r + i \omega_i$,
the so-called quasinormal frequencies. 

In theory, they are the outcome of a spreading wave through a
gravitational potential, for which outgoing waves are the proper
boundary conditions in terms of the tortoise coordinate, dispersing
away of the potential barrier (since 
nothing comes out of the horizon). In general terms these conditions
are written as a plane wave field $\Psi$ in the limits of a coordinate
$x$ as $\Psi \Big|_{x_{\pm \infty}} \rightarrow e^{\mp i\omega x}$. 

The typical vibrational spectrum that emerges from such spreading
is determined by two sets of features, the geometry
parameters and the inner field characteristics. Related to these
characteristics we evidence two useful aspects in this letter, namely,
the angular momentum (or equivalent) and overtone number. Those
quantities are natural numbers connected, respectively, to the angular
part of the motion equation and to the label of a quantized wave
solution of its radial part.  

These numbers establish a very special feature in the
spectrum, whenever they are high, we have fixed values of 'density
$\omega$'s\footnote{Fixed $\omega_r /\ell$ and $\omega_i / \nu$.}. To
exemplify let us recall the result in Ref. \cite{Cardoso_2009},
$\omega = \ell \Omega_c - i\nu|\lambda |,$ which established the
equivalence of the real and imaginary parts of $\omega  /\alpha$,
$\alpha\rightarrow \ell , \nu$, with the geodesic angular velocity,
$\Omega_c$ and Lyapunov exponent, $\lambda$.  

Such astonishing result dictates a family of solutions known as the
photon sphere quasinormal modes\footnote{Other families not related
  to those may as well be present.  We take as an example, the near
  extremal, the cosmological, \cite{Cardoso_2018, Destounis_2019} and
  the acceleration families of modes
  \cite{destounis2020accelerating}}, which keep a close relation to
the outermost photon orbit around the black hole (proved to be
unstable). These modes are obtained traditionally with the WKB method
(more details in the next section). 

The relation between quasinormal modes and geodesic quantities
reported in \cite{Cardoso_2009} were recently revived connecting it
with black holes shadows as the one reported by the Event Horizon
Telescope last year \cite{Akiyama:2019cqa, Akiyama:2019eap}. In the
same way, gravitational lensing observables may be strictly connected
to the perturbed solution, {\it viz.} to those oscillations,  as pointed out in \cite{Stefanov_2010}.

As a general feature in spherically symmetric spacetimes the photon
sphere region collapses to the value of the maxima of every black hole
potential no matter which field is being considered. Such a result
brings new light in the shadow phenomenon and its connection to
perturbations as we will further see along this work.  

This letter is organized as follows, in section \ref{sec.2} we
demonstrate the equivalence of quasinormal modes at the eikonal limit
and black hole shadow (first conjectured in \cite{Jusufi:2019ltj})
following with examples of the identification in section
\ref{sec.3}. We summarize our discussion in section \ref{sec.4}.

\section{Circular photon orbit, shadows, and quasinormal modes via WKB}\label{sec.2}

Let us begin with a sufficiently generic line element that could
represent several D-dimensional black holes with spherical symmetry given by
\be
\label{e1}
ds^2=-fdt^2 + \frac{dr^2}{f} + r^2 d \Omega_{D-2}^2.
\ee
Here the spherical symmetry implies $f=f(r)$. Many different field motion equations as well as a linear gravitational perturbation can be expressed with a master formula written as
\be
\label{e2}
\left[ \frac{\partial^2}{\partial x^2}-\frac{\partial^2}{\partial t^2} + V(r) \right] \Psi (r,t) = 0
\ee
where $x$ is a typical tortoise radial coordinate which maps its
infinities (of the physically field propagating) into the singular
points of $r$ ({\it i.e.}, asymptotical infinities or horizons) via
$dx=f^{-1}dr$. The potential in Eq. (\ref{e2}) can be generically
expressed as a centrifugal term plus a function of the radial
coordinate. This function encodes all possible information about the
geometry of the spacetime, the theory under consideration ({\it e.g.},
general relativity, Gauss-Bonnet, Horndeski, etc.), and the type of
propagating field. It can be written as
\be
\label{e3}
V(r)= f(r) \left( g(r) + \frac{\ell (\ell +D-3)}{r^2}\right)
\ee
for the purpose of use in WKB method. Now, for the same spacetime defined in (\ref{e1}), the limiting unstable photon orbit can be defined as the solution of the equation~\cite{Perlick:2015vta}
\be
\label{e4}
\frac{d}{dr}\left( \frac{r^2}{f(r)} \right)\Bigg|_{r=r_{ps}}=0
\ee
Such equation contemplates a multitude of black hole solutions in
general relativity, {\it e.g.} solutions with mass, charge, cosmological constant (positive), and anisotropic fluids.
The concept of limiting orbit not only relates to the last stable photon geodesics around the hole, but also defines the idea of {\it cone of avoidance} \cite{Chandrasekhar:1985kt} as the region whose angle represents a 'dark place in the sky' seen by a 'looking-backwards-observer' falling into the hole, 
\be
\label{e5}
\tan \frac{\Theta}{2} = \sqrt{r^2 f} \frac{d\phi}{d r}.
\ee
In the Schwarzschild geometry, for instance, $\tan \frac{\Theta}{2}
\propto (r- r_{ps})^{-1}(r-r_h)^{1/2}$, and at the
point $r = r_{ps}$ this observer has a cone of avoidance of exactly $\pi$ (no closed stable geodesic for $r<r_{ps}$). 

Related to the same limit we stress another important quantity, the shadow radius of a black hole \cite{Jusufi:2019ltj,Jusufi:2020dhz,Bisnovatyi-Kogan:2017kii,Perlick:2015vta}, defined in terms of the photon sphere unstable orbit as
\be
\label{e6}
\textgoth{S}= \frac{r}{\sqrt{f}}\Bigg|_{r=r_{ps}}\,,
\ee
which corresponds to the angular semi-diameter of the shadow around a black hole as seen by a distant observer.

The main goal of this work is to provide the missing link that establishes the correspondence of the real part of the quasinormal modes (for whatever kind of perturbations) at the eikonal limit and the inverse of the shadow radius $\textgoth{S}$ of the black hole, and present examples of it.

The quasinormal modes were studied with a multitude of methods along the last decades. For an extensive review refer to \cite{Konoplya:2011qq}. Here we employ one of these methods, the semi-analytical WKB approximation, whose application in gravitational theory was first shown in the 80's \cite{Iyer:1986np,Kokkotas:1988fm,Seidel:1989bp}. The method was nicely extended to $6^{th}$ order \cite{Konoplya_2003}, and in 2017 to $13^{th}$ order \cite{Matyjasek:2017psv}.

For the purpose of our work the $3^{rd}$ order expansion reads 
\be
\nonumber
\omega = \Bigg\{  V + \frac{V_4}{8V_2}\left( \nu^2+\frac{1}{4}\right) - \left( \frac{7 + 60\nu^2}{288}\right) \frac{V_3^2}{V_2^2}+i\nu \sqrt{-2V_2}\left[\frac{1}{2V_2}\left[\frac{5V_3^4(77+188\nu^2)}{6912V_2^4} \right. \right. \\
\nonumber
\left. \left. -\frac{V_3^2V_4(51+100\nu^2 )}{384V_2^3}+ \frac{V_4^2(67+68\nu^2 )}{2304V_2^2}+\frac{V_5V_3(19+28\nu^2 )}{288V_2^2}+\frac{V_6(5+4\nu^2 )}{288V_2}\right] -1 \right]  \Bigg\}^{1/2}_{r=r_0}\\
\label{e7}
\ee
which produces the same expansion for the first terms of the eikonal
limit when compared to the $4^{th}$ to $6^{th}$ order representation. Here $V_i$ represents the {\it i-th} derivative of the potential $V$ and $\nu = n+\frac{1}{2}$, $n \in \mathbb{N}$, is the overtone number.  The above expression for $\omega$ is to be taken at the point $r_0$, defined as the maximum value of the potential $V$ through 
\be
\label{e8}
\frac{dV}{dr}\Bigg|_{r=r_0}= \left[ \ell (\ell +D-3)\frac{d}{dr}\left(\frac{f(r)}{r^2} \right)+ \frac{d}{dr}(f(r)g(r)) \right] \Bigg|_{r=r_0}= 0.
\ee
The latter equation renders different values of $r_0$ depending on
the physical field, theory, and black hole (expressed through $g$),
but to leading - and first sub-leading - order its solution at the eikonal limit is the very simple relation, 
\be
\label{e9}
\frac{d}{dr}\left(\frac{f(r)}{r^2}\right)  \Bigg|_{r=r_0}= 0.
\ee
The interesting fact is that $r_0$ and $r_{ps}$ represent the same point, defining $G = \frac{f(r)}{r^2}$, those equations can be written in the form
\be
\label{ex1}
\frac{d}{dr}\left(G\right)  \Bigg|_{r=r_0}= 0, \hspace{2.0cm} \frac{d}{dr}\left(G^{-1}\right)  \Bigg|_{r=r_{ps}}= 0.
\ee
As a consequence, $r_{ps}=r_0$, as long as $G^{-2}|_{r=r_{ps}} \neq 0$, which is the case in general.

 This result states that for every spherically symmetric black hole that possesses a photon sphere, the position of the maximum of the potential of motion equations of fields corresponds to the stability threshold for the circular null geodesic around the structure. 

Finally, by expanding the relation (\ref{e7}), we obtain at the eikonal regime, 
\be
\nonumber
\omega = \omega_R - i \omega_I \rightarrow \left[\ell \frac{\sqrt{f(r)}}{r}\Bigg|_{r=r_0} +  \frac{(D-3)\sqrt{f(r)}}{2r}\Bigg|_{r=r_0} + \mathcal{O}(\ell^{-1})\right]_R -\\
 i \left[\frac{\nu}{\sqrt{2}}\frac{\sqrt{f(r)}}{r}\Bigg|_{r=r_0} 
\sqrt{6rf'-6f-r^2f''-r^2f^{-1}{f'}^2}\Bigg|_{r=r_0} +  \mathcal{O}(\ell^{-1})\right]_I.
\label{e10}
\ee 
The imaginary part of the approximation to leading order reads
\be
\label{e11}
 \omega_I = \frac{2\nu +1}{2\sqrt{2}}\textgoth{S}^{-1}\sqrt{2f-r^2f''} +\mathcal{O}(\ell^{-1}),
\ee
in which the square root term is the second derivative of the potential at its maximum (multiplied by other constants), a harmonic oscillator related term. 
As for the real part, it corresponds - to leading order  - exactly to the shadow radius of the black  hole and to sub-leading regime to half of its value,
\be
\label{e12}
\omega_R = \textgoth{S}^{-1}\left(\ell + \frac{D-3}{2} +\mathcal{O}(\ell^{-1})\right).
\ee
As conjectured in \cite{Jusufi:2019ltj} and here demonstrated, the result has an interesting interpretation for the real part of
the quasinormal modes at high angular momentum regime as the shadow
radius observed in black holes with spherical symmetry. The
identification first appeared in \cite{Jusufi:2019ltj} and was 
further investigated for rotating spacetimes in
\cite{Jusufi:2020dhz}. In what follows we will give some examples of its application in known black hole systems.

\section{Results}\label{sec.3}

In this section we apply the identification of the
real part of quasinormal modes at the eikonal limit with the radius of
black hole shadow for two different families of black hole
solutions, the D-dimensional Tangherlini metric and a black hole
surrounded by anisotropic fluids in 4 dimensions.

\subsection{D-dimensional Tangherlini black hole}

The metric corresponding to D-dimensional Tangherlini black
hole~\cite{Tangherlini:1963bw} has the same form as Eq.(\ref{e1}), 
where the metric function $f(r)$ and the angular part $d\Omega_{D-2}
^2$ are given by 
\begin{eqnarray}\label{tang2}
f(r) &=& 1-\frac{\mu}{r^{D-3}} \nonumber \\
d\Omega_{D-2} ^2 &=& \sum_{i=1} ^{D-2} \left( \prod_{n=1} ^{i-1}
\sin^2 \theta_n \right) d\theta_i ^2 \,.
\end{eqnarray}
The parameter $\mu$ is related to the mass $M$ of the black hole as
\begin{equation}
\mu = \frac{16 \pi M}{(D-2) \Omega_{D-2}}\,, \hbox{ with } \;
\Omega_{D-2}=\frac{2\pi^{(D-1)/2}}{\Gamma\left(\frac{D-1}{2}\right)} \,.
\end{equation}

In order to find the radius of the photon sphere we will use the usual
Lagrangian formalism which for null geodesics gives
\begin{equation}
{\cal L} = \frac{1}{2} g_{\mu\nu} \dot x ^\mu \dot x ^\nu = 0 \,.
\end{equation}
Finding the canonically conjugated momenta and substituting back into
this Lagrangian we can decouple the angular part and obtain the radial
equation for a photon geodesic in the form,
\begin{equation}\label{req}
\dot r ^2 + V_T(r) = 0\,,
\end{equation}
where the potential $V_T(r)$ can be written as
\begin{equation}\label{VT}
V_T(r) = \frac{f(r)}{r^2} (K^2 + L^2) - E^2\,.
\end{equation}
Here $E$ and $L$ are the constants of motion associated to $t$ and
$\theta_2$ coordinates (energy and angular momentum, respectively) and
$K$ is a decoupling constant~\cite{PhysRev.174.1559}. Notice that we set $\theta_1=\pi/2$ as
usual. 

Applying the photon sphere conditions,
\begin{equation}\label{rpscond}
V_T(r_{ps})=0\,, \quad \left. \frac{dV_T}{dr}\right\vert_{r=r_{ps}} =0\,, \quad
\left. \frac{d^2V_T}{dr^2}\right\vert_{r=r_{ps}} <0\,,
\end{equation}
we obtain
\begin{equation}
r_{ps} = \left[\frac{8\pi M (D-1)}{\Omega_{D-2} (D-2)}\right]^{\frac{1}{D-3}}\,,
\end{equation}
together with a relation between the constants of motion,
\begin{equation}
E^2 = \frac{(D-3) (K+L^2)}{(D-1)r_{ps} ^2} \,.
\end{equation}
Thus, the radius of the black hole shadow becomes
\begin{equation}\label{mysh}
\textgoth{S} = \frac{r_{ps}}{\sqrt{f(r_{ps})}} = \sqrt{\frac{D-1}{D-3}}\, r_{ps}\,.
\end{equation}
A similar result was found in~\cite{Singh_2018} using a different
method.

For the other side of the correspondence in Eq.(\ref{e12}) we can
consider the massive scalar perturbation potential for a Tangherlini black
hole given by~\cite{Zhidenko:2006rs}
\begin{equation}
V_s(r) = \frac{(D-2)(D-4)}{4r^2} f^2(r) + \frac{(D-2)}{2r} f'(r) f(r) + \left[
  \frac{\ell (\ell+D-3)}{r^2} + m^2 \right] f(r)\,, 
\end{equation}
where $m$ represents the mass of the scalar perturbation. We applied
the 6$^{th}$ order WKB method in order to obtain the quasinormal
frequencies for the fundamental mode. 

In Tables \ref{TD4}--\ref{TD10} we show these frequencies 
for different dimensions, perturbation masses, and multipole
numbers. The last two lines correspond to the frequencies 
obtained from Eq.(\ref{e12}) using the shadow approach to leading and
to first subleading order, {\it i.e.}, $\omega_0=\ell/\textgoth{S}$ and
$\omega_1=[\ell+(D-3)/2]/\textgoth{S}$, respectively, with
$\textgoth{S}$ given by (\ref{mysh}). By comparing the 
frequencies in these tables we see that the conjecture in
Eq.(\ref{e12}) is fulfilled as we reach the eikonal limit. Moreover,
the scalar perturbation mass does not affect the results in this same
limit. 

\begin{table}[!htb]
\footnotesize
\begin{center}
\begin{tabular}{|c|c|c|c|c|}
\hline
$D=4$ & $\ell=10$ & $\ell=10^2$ & $\ell=10^3$ & $\ell=10^5$ \\ \hline
\hline
$m=0.0$ & 2.026568929 & 19.34186945 & 192.5463789 & 19245.10520 \\ \hline
$m=0.3$ & 2.033933121 & 19.34264493 & 192.5464569 & 19245.10520 \\ \hline
$m=0.6$ & 2.056095784 & 19.34497141 & 192.5466905 & 19245.10520 \\ \hline
$m=0.9$ & 2.093217069 & 19.34884911 & 192.5470800 & 19245.10520 \\ \hline \hline
$\omega_0$ & 1.924500898 & 19.24500898 & 192.4500898 & 19245.00898 \\ \hline
$\omega_1$ & 2.020725943 & 19.34123403 & 192.5463149 & 19245.10521 \\ \hline
\end{tabular}
\caption{Real part of quasinormal frequencies $\omega_R$ for different
multipole numbers $\ell$ and scalar perturbation masses $m$ in $D=4$
Tangherlini black hole.}
\label{TD4}
\end{center}
\end{table} 

\begin{table}[!htb]
\footnotesize
\begin{center}
\begin{tabular}{|c|c|c|c|c|}
\hline
$D=5$ & $\ell=10$ & $\ell=10^2$ & $\ell=10^3$ & $\ell=10^5$ \\ \hline
\hline
$m=0.0$ & 6.001713288 & 54.81646560 & 543.2440142 & 54270.63680 \\ \hline
$m=0.3$ & 6.005426726 & 54.81687599 & 543.2440558 & 54270.63681 \\ \hline
$m=0.6$ & 6.016566980 & 54.81810723 & 543.2441800 & 54270.63681 \\ \hline
$m=0.9$ & 6.035135991 & 54.82015928 & 543.2443871 & 54270.63681 \\ \hline \hline
$\omega_0$ & 5.427009412 & 54.27009412 & 542.7009412 & 54270.09412 \\ \hline
$\omega_1$ & 5.969710354 & 54.81279507 & 543.2436422 & 54270.63683 \\ \hline
\end{tabular}
\caption{Real part of quasinormal frequencies $\omega_R$ for different
multipole numbers $\ell$ and scalar perturbation masses $m$ in $D=5$
Tangherlini black hole.}
\label{TD5}
\end{center}
\end{table}

\begin{table}[!htb]
\footnotesize
\begin{center}
\begin{tabular}{|c|c|c|c|c|}
\hline
$D=10$ & $\ell=10$ & $\ell=10^2$ & $\ell=10^3$ & $\ell=10^5$ \\ \hline
\hline
$m=0.0$ & 12.15525866 & 91.93763045 & 891.1897864 & 88811.03578 \\ \hline
$m=0.3$ & 12.15807589 & 91.93801097 & 891.1898256 & 88811.03578 \\ \hline
$m=0.6$ & 12.16652560 & 91.93915253 & 891.1899435 & 88811.03578 \\ \hline
$m=0.9$ & 12.18060185 & 91.94105512 & 891.1901398 & 88811.03578 \\ \hline \hline
$\omega_0$ & 8.880792741 & 88.80792741 & 888.0792741 & 88807.92741 \\ \hline
$\omega_1$ & 11.98907021 & 91.91620487 & 891.1875518 & 88811.03569 \\ \hline
\end{tabular}
\caption{Real part of quasinormal frequencies $\omega_R$ for different
multipole numbers $\ell$ and scalar perturbation masses $m$ in $D=10$
Tangherlini black hole.}
\label{TD10}
\end{center}
\end{table} 


\subsection{Black holes surrounded by anisotropic fluids} 
 
The line element describing the geometry of a spherically symmetric
black hole surrounded by an anisotropic fluid \cite{Kiselev:2002dx} is the same as in Eq.(\ref{e1}) 
with
\begin{equation}\label{gtt}
f(r)=1-\frac{2M}{r}+\frac{Q^2}{r^2}-\frac{c}{r^{3w_f+1}},
\end{equation}
characterized by the black hole mass $M$, charge $Q$, the parameter
$w_{f}$ obeying the equation of state $p=w_f\rho$ (being $p$ and
$\rho$ the pressure and energy density of the fluid, respectively) of the anisotropic fluid and $c$ is a dimensional normalization constant related to the presence of surrounding fluid.

It is worthwhile to mention some special cases of the solution
(\ref{gtt}). The Schwarzschild solution is recovered in two cases,
for $Q=c=0$ and for $Q=w_f=0$ having its mass shifted to $2M-c$.
For $Q=0$ and $w_f=-1$ we have the Schwarzschild-(anti) de Sitter
black hole with $3c$ playing the role of a cosmological constant. The
charged case includes the Reissner-Nordstr\"om-de Sitter solution for
$w_f=-1$.

For our discussion we will consider two representative cases
of (\ref{gtt}), namely, $w_f=-1/2$ and $w_f=-2/3$. For a wide range
of parameters those cases admit three horizons, an inner Cauchy
horizon  at $r=r_{-}$, the event horizon $r=r_{+}$, and a
cosmological-like horizon $r=r_c$. The full discussion of causal
structure of such geometries is explored in
\cite{Cuadros-Melgar:2020shz}. 

The equation that determines the circular photon orbit, as shown in
Tangherlini case, is obtained through the value $r=r_{ps}$ that turns
the effective potential for the photon a maximum as expressed by
conditions (\ref{rpscond}), where in this case
\begin{equation}\label{pot_foton}
V (r)=\frac{L^{2}}{r^2}f(r),
\end{equation}
with $L$ standing for the angular momentum of the particle.
Thus, for the case under consideration here the equation for the photon orbit is given by
\begin{equation}\label{orbita_foton}
3c(1+w_f)r_{ps}^{(1-3w_f)}-2r_{ps}^{2}+6Mr_{ps}-4Q^{2}=0\,.
\end{equation}
Notice that the solution of this equation depends crucially on the
fluid nature encoded by the parameter $w_f$. 

Following the recipe outlined in the Sec.\ref{sec.2}, we first obtain
the radius of circular photon orbit $r=r_{ps}$ for a given $w_f$ using
the equation (\ref{orbita_foton}) and, then, substituting back into the expression (\ref{e12}) together with (\ref{e6}) we have the real part of quasinormal frequencies $\omega_R$ at the eikonal limit,
\begin{equation}\label{frequencias_eikonal}
\frac{\omega_R}{\left(\ell+\frac{1}{2}\right)}=\frac{1}{\textgoth{S}}=\frac{1}{r_{ps}}\sqrt{1-\frac{2M}{r_{ps}}+\frac{Q^2}{r_{ps}^{2}}-\frac{c}{r_{ps}^{3w_f+1}}}\,.
\end{equation}

\begin{table}[htbp]
\footnotesize
\begin{center}
\begin{tabular}{|c|c|c||c|c|}
\hline
\multicolumn{1}{c}{} & \multicolumn{2}{c}{$w_f=-1/2$} & \multicolumn{2}{c}{$w_f=-2/3$} \\
\cline{1-5}
$c/c_{max}$ & \multicolumn{1}{c|}{$(\ell+1/2)/\omega_R$} & \multicolumn{1}{c||}{$\textgoth{S}$} & $(\ell+1/2)/\omega_R$& $\textgoth{S}$ \\
\hline
0.1 & $5.354485444$ & $5.354485444$ & 5.271473269 & 5.271473269 \\
\hline
0.3 & $6.377550833$ & $6.377550833$ & 6.063015498 & 6.063015499  \\
\hline
0.5 & $7.992925474$ & $7.992925474$ & 7.288469762& 7.288469762  \\
\hline
0.7 & $11.07073257$ & $11.07073257$ & 9.580273496& 9.580273495  \\
\hline
0.9 & $21.05924133$ & $21.05924133$ & 16.94705693 & 16.94705693  \\
\hline
\end{tabular}
\caption{Comparison between $(\ell+1/2)/\omega_R$ and the shadow radius $\textgoth{S}$ for $\ell=10^{5}$, $M=2Q=1$, $c_{max}\approx 0.2751$ for $w_f=-1/2$ and $c_{max}\approx 0.1292$ for $w_f=-2/3$.}
\label{tab1}
\end{center}
\end{table}

In Table \ref{tab1} we show the dependence of
$(\ell+\frac{1}{2})/\omega_R$ and the shadow radius $\textgoth{S}$
with the parameter $c$ in the cases $w_f=-1/2$ and $w_f=-2/3$. Notice that in the first column the
parameter $c$ is normalized by $c_{max}$, which is the maximum value
permitted for $c$ in order to avoid naked singularities
\cite{Cuadros-Melgar:2020shz}. From those results we observe that as
the parameter $c$ of the anisotropic fluid increases, the radius of the black hole shadow $\textgoth{S}$ gets bigger in comparison to the case in the absence of the fluid. This result is similar to that in the case of the Schwarzschild black hole surrounded by a homogeneous plasma acting as a dispersive medium for the light rays \cite{Bisnovatyi-Kogan:2017kii}. Also, we observe that the correspondence between $(\ell+\frac{1}{2})/\omega_R$ and the shadow radius $\textgoth{S}$ at the eikonal limit is fulfilled in this case as well.  

In Table \ref{tab2} we present the behavior of $\omega_R$ as we increase the multipole number $\ell$ towards the eikonal limit for the case $w_f=-1/2$. A similar qualitative result is obtained for $w_f=-2/3$. 
\begin{table}[htbp]
\footnotesize
\begin{center}
\begin{tabular}{|c|c|c|c|c|}
\hline
$c/c_{max}$ & \multicolumn{1}{c|}{$\ell=10$} & \multicolumn{1}{c|}{$\ell=10^{2}$} & $\ell=10^{3}$& $\ell=10^{5}$\\
\hline
0.1 & $1.9613129$ & $18.769346$ & 186.85269 & 18676.02425 \\
\hline
0.3 & $1.6462219$ & $15.758382$ & 156.87840 & 15680.07886  \\
\hline
0.5 & $1.31311184$ & $12.5735616$ & 125.173187& 12511.12629 \\
\hline
0.7 & $0.94771582$ & $9.0779137$ & 90.373416& 12511.126286   \\
\hline
0.9 & $0.49800157$ & $4.7721899$ & 47.508828 & 9032.871070  \\
\hline
\end{tabular}
\caption{Real part of quasinormal frequencies $\omega_R$ as $\ell$
  increases with $w_f=-1/2$, $M=2Q=1$, and $c_{max}\approx 0.2751$.}
\label{tab2}
\end{center}
\end{table}

\begin{figure}[htb]
\begin{center}
\includegraphics[height=5.3cm, width=8cm]{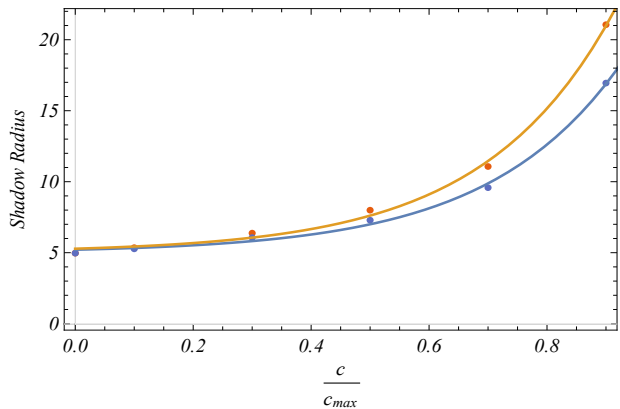}
\caption{Shadow radius as a function of the anisotropic fluid
  parameter $c/c_{max}$. The upper fitting curve (orange) refers to
  the case $w_f=-1/2$ and the bottom one (blue) to $w_f=-2/3$. In both
  curves we have set $M=2Q=1$.}
\label{graphs_fluid}
\end{center}
\end{figure}
In Fig.(\ref{graphs_fluid}) we show the shadow radius for several values of $c/c_{max}$ and the corresponding fitting curves for each case of interest $w_f=-1/2$ and $w_f=-2/3$. For small values of $c/c_{max}$ the shadow radius does not depend strongly on $w_f$. However, a different picture comes up as $c/c_{max}$ increases showing very different values depending on the fluid characteristics.

\section{Discussion}\label{sec.4}

In this paper we found an analytical relation between the eikonal
limit of quasinormal frequencies and the black hole shadow, a 
result first conjectured in~\cite{Jusufi:2019ltj}.
We show that every spherically symmetric black hole having an
outermost photon orbit has an identification of the maxima of the
potentials  corresponding to perturbation fields and null geodesics.

In order to illustrate the correspondence we compute the quasinormal
frequencies and shadow radius for two families of black holes, {\it
  i.e.}, Tangherlini and a black hole surrounded by anisotropic fluids,
verifying its validity. 

Further investigation includes the computation of quasinormal modes in
more realistic scenarios like black holes (or other astrophysical
objects) with accretion disks or surrounded by plasmas.



\section*{References}

\bibliography{referencias}

\begin{thebibliography}{24}%
\makeatletter
\providecommand \@ifxundefined [1]{%
 \@ifx{#1\undefined}
}%
\providecommand \@ifnum [1]{%
 \ifnum #1\expandafter \@firstoftwo
 \else \expandafter \@secondoftwo
 \fi
}%
\providecommand \@ifx [1]{%
 \ifx #1\expandafter \@firstoftwo
 \else \expandafter \@secondoftwo
 \fi
}%
\providecommand \natexlab [1]{#1}%
\providecommand \enquote  [1]{``#1''}%
\providecommand \bibnamefont  [1]{#1}%
\providecommand \bibfnamefont [1]{#1}%
\providecommand \citenamefont [1]{#1}%
\providecommand \href@noop [0]{\@secondoftwo}%
\providecommand \href [0]{\begingroup \@sanitize@url \@href}%
\providecommand \@href[1]{\@@startlink{#1}\@@href}%
\providecommand \@@href[1]{\endgroup#1\@@endlink}%
\providecommand \@sanitize@url [0]{\catcode `\\12\catcode `\$12\catcode
  `\&12\catcode `\#12\catcode `\^12\catcode `\_12\catcode `\%12\relax}%
\providecommand \@@startlink[1]{}%
\providecommand \@@endlink[0]{}%
\providecommand \url  [0]{\begingroup\@sanitize@url \@url }%
\providecommand \@url [1]{\endgroup\@href {#1}{\urlprefix }}%
\providecommand \urlprefix  [0]{URL }%
\providecommand \Eprint [0]{\href }%
\providecommand \doibase [0]{http://dx.doi.org/}%
\providecommand \selectlanguage [0]{\@gobble}%
\providecommand \bibinfo  [0]{\@secondoftwo}%
\providecommand \bibfield  [0]{\@secondoftwo}%
\providecommand \translation [1]{[#1]}%
\providecommand \BibitemOpen [0]{}%
\providecommand \bibitemStop [0]{}%
\providecommand \bibitemNoStop [0]{.\EOS\space}%
\providecommand \EOS [0]{\spacefactor3000\relax}%
\providecommand \BibitemShut  [1]{\csname bibitem#1\endcsname}%
\let\auto@bib@innerbib\@empty
\bibitem [{\citenamefont {Cardoso}\ \emph {et~al.}(2009)\citenamefont
  {Cardoso}, \citenamefont {Miranda}, \citenamefont {Berti}, \citenamefont
  {Witek},\ and\ \citenamefont {Zanchin}}]{Cardoso_2009}%
  \BibitemOpen
  \bibfield  {author} {\bibinfo {author} {\bibfnamefont {V.}~\bibnamefont
  {Cardoso}}, \bibinfo {author} {\bibfnamefont {A.~S.}\ \bibnamefont
  {Miranda}}, \bibinfo {author} {\bibfnamefont {E.}~\bibnamefont {Berti}},
  \bibinfo {author} {\bibfnamefont {H.}~\bibnamefont {Witek}}, \ and\ \bibinfo
  {author} {\bibfnamefont {V.~T.}\ \bibnamefont {Zanchin}},\ }\href {\doibase
  10.1103/PhysRevD.79.064016} {\bibfield  {journal} {\bibinfo  {journal} {Phys.
  Rev. D}\ }\textbf {\bibinfo {volume} {79}},\ \bibinfo {pages} {064016}
  (\bibinfo {year} {2009})},\ \Eprint {http://arxiv.org/abs/0812.1806}
  {arXiv:0812.1806 [hep-th]} \BibitemShut {NoStop}%
\bibitem [{\citenamefont {Cardoso}\ \emph {et~al.}(2018)\citenamefont
  {Cardoso}, \citenamefont {Costa}, \citenamefont {Destounis}, \citenamefont
  {Hintz},\ and\ \citenamefont {Jansen}}]{Cardoso_2018}%
  \BibitemOpen
  \bibfield  {author} {\bibinfo {author} {\bibfnamefont {V.}~\bibnamefont
  {Cardoso}}, \bibinfo {author} {\bibfnamefont {J.~L.}\ \bibnamefont {Costa}},
  \bibinfo {author} {\bibfnamefont {K.}~\bibnamefont {Destounis}}, \bibinfo
  {author} {\bibfnamefont {P.}~\bibnamefont {Hintz}}, \ and\ \bibinfo {author}
  {\bibfnamefont {A.}~\bibnamefont {Jansen}},\ }\href {\doibase
  10.1103/PhysRevLett.120.031103} {\bibfield  {journal} {\bibinfo  {journal}
  {Phys. Rev. Lett.}\ }\textbf {\bibinfo {volume} {120}},\ \bibinfo {pages}
  {031103} (\bibinfo {year} {2018})},\ \Eprint
  {http://arxiv.org/abs/1711.10502} {arXiv:1711.10502 [gr-qc]} \BibitemShut
  {NoStop}%
\bibitem [{\citenamefont {Destounis}\ \emph {et~al.}(2019)\citenamefont
  {Destounis}, \citenamefont {Fontana}, \citenamefont {Mena},\ and\
  \citenamefont {Papantonopoulos}}]{Destounis_2019}%
  \BibitemOpen
  \bibfield  {author} {\bibinfo {author} {\bibfnamefont {K.}~\bibnamefont
  {Destounis}}, \bibinfo {author} {\bibfnamefont {R.~D.}\ \bibnamefont
  {Fontana}}, \bibinfo {author} {\bibfnamefont {F.~C.}\ \bibnamefont {Mena}}, \
  and\ \bibinfo {author} {\bibfnamefont {E.}~\bibnamefont {Papantonopoulos}},\
  }\href {\doibase 10.1007/JHEP10(2019)280} {\bibfield  {journal} {\bibinfo
  {journal} {JHEP}\ }\textbf {\bibinfo {volume} {10}},\ \bibinfo {pages} {280}
  (\bibinfo {year} {2019})},\ \Eprint {http://arxiv.org/abs/1908.09842}
  {arXiv:1908.09842 [gr-qc]} \BibitemShut {NoStop}%
\bibitem [{\citenamefont {Destounis}\ \emph {et~al.}(2020)\citenamefont
  {Destounis}, \citenamefont {Fontana},\ and\ \citenamefont
  {Mena}}]{destounis2020accelerating}%
  \BibitemOpen
  \bibfield  {author} {\bibinfo {author} {\bibfnamefont {K.}~\bibnamefont
  {Destounis}}, \bibinfo {author} {\bibfnamefont {R.~D.~B.}\ \bibnamefont
  {Fontana}}, \ and\ \bibinfo {author} {\bibfnamefont {F.~C.}\ \bibnamefont
  {Mena}},\ }\href@noop {} {\  (\bibinfo {year} {2020})},\ \Eprint
  {http://arxiv.org/abs/2005.03028} {arXiv:2005.03028 [gr-qc]} \BibitemShut
  {NoStop}%
\bibitem [{\citenamefont {Akiyama}\ \emph
  {et~al.}(2019{\natexlab{a}})\citenamefont {Akiyama} \emph
  {et~al.}}]{Akiyama:2019cqa}%
  \BibitemOpen
  \bibfield  {author} {\bibinfo {author} {\bibfnamefont {K.}~\bibnamefont
  {Akiyama}} \emph {et~al.} (\bibinfo {collaboration} {Event Horizon
  Telescope}),\ }\href {\doibase 10.3847/2041-8213/ab0ec7} {\bibfield
  {journal} {\bibinfo  {journal} {Astrophys. J.}\ }\textbf {\bibinfo {volume}
  {875}},\ \bibinfo {pages} {L1} (\bibinfo {year} {2019}{\natexlab{a}})},\
  \Eprint {http://arxiv.org/abs/1906.11238} {arXiv:1906.11238 [astro-ph.GA]}
  \BibitemShut {NoStop}%
\bibitem [{\citenamefont {Akiyama}\ \emph
  {et~al.}(2019{\natexlab{b}})\citenamefont {Akiyama} \emph
  {et~al.}}]{Akiyama:2019eap}%
  \BibitemOpen
  \bibfield  {author} {\bibinfo {author} {\bibfnamefont {K.}~\bibnamefont
  {Akiyama}} \emph {et~al.} (\bibinfo {collaboration} {Event Horizon
  Telescope}),\ }\href {\doibase 10.3847/2041-8213/ab1141} {\bibfield
  {journal} {\bibinfo  {journal} {Astrophys. J.}\ }\textbf {\bibinfo {volume}
  {875}},\ \bibinfo {pages} {L6} (\bibinfo {year} {2019}{\natexlab{b}})},\
  \Eprint {http://arxiv.org/abs/1906.11243} {arXiv:1906.11243 [astro-ph.GA]}
  \BibitemShut {NoStop}%
\bibitem [{\citenamefont {Stefanov}\ \emph {et~al.}(2010)\citenamefont
  {Stefanov}, \citenamefont {Yazadjiev},\ and\ \citenamefont
  {Gyulchev}}]{Stefanov_2010}%
  \BibitemOpen
  \bibfield  {author} {\bibinfo {author} {\bibfnamefont {I.~Z.}\ \bibnamefont
  {Stefanov}}, \bibinfo {author} {\bibfnamefont {S.~S.}\ \bibnamefont
  {Yazadjiev}}, \ and\ \bibinfo {author} {\bibfnamefont {G.~G.}\ \bibnamefont
  {Gyulchev}},\ }\href {\doibase 10.1103/PhysRevLett.104.251103} {\bibfield
  {journal} {\bibinfo  {journal} {Phys. Rev. Lett.}\ }\textbf {\bibinfo
  {volume} {104}},\ \bibinfo {pages} {251103} (\bibinfo {year} {2010})},\
  \Eprint {http://arxiv.org/abs/1003.1609} {arXiv:1003.1609 [gr-qc]}
  \BibitemShut {NoStop}%
\bibitem [{\citenamefont {Jusufi}(2020{\natexlab{a}})}]{Jusufi:2019ltj}%
  \BibitemOpen
  \bibfield  {author} {\bibinfo {author} {\bibfnamefont {K.}~\bibnamefont
  {Jusufi}},\ }\href {\doibase 10.1103/PhysRevD.101.084055} {\bibfield
  {journal} {\bibinfo  {journal} {Phys. Rev. D}\ }\textbf {\bibinfo {volume}
  {101}},\ \bibinfo {pages} {084055} (\bibinfo {year} {2020}{\natexlab{a}})},\
  \Eprint {http://arxiv.org/abs/1912.13320} {arXiv:1912.13320 [gr-qc]}
  \BibitemShut {NoStop}%
\bibitem [{\citenamefont {Perlick}\ \emph {et~al.}(2015)\citenamefont
  {Perlick}, \citenamefont {Tsupko},\ and\ \citenamefont
  {Bisnovatyi-Kogan}}]{Perlick:2015vta}%
  \BibitemOpen
  \bibfield  {author} {\bibinfo {author} {\bibfnamefont {V.}~\bibnamefont
  {Perlick}}, \bibinfo {author} {\bibfnamefont {O.~Y.}\ \bibnamefont {Tsupko}},
  \ and\ \bibinfo {author} {\bibfnamefont {G.~S.}\ \bibnamefont
  {Bisnovatyi-Kogan}},\ }\href {\doibase 10.1103/PhysRevD.92.104031} {\bibfield
   {journal} {\bibinfo  {journal} {Phys. Rev. D}\ }\textbf {\bibinfo {volume}
  {92}},\ \bibinfo {pages} {104031} (\bibinfo {year} {2015})},\ \Eprint
  {http://arxiv.org/abs/1507.04217} {arXiv:1507.04217 [gr-qc]} \BibitemShut
  {NoStop}%
\bibitem [{\citenamefont {Chandrasekhar}(1985)}]{Chandrasekhar:1985kt}%
  \BibitemOpen
  \bibfield  {author} {\bibinfo {author} {\bibfnamefont {S.}~\bibnamefont
  {Chandrasekhar}},\ }\href@noop {} {\emph {\bibinfo {title} {{The mathematical
  theory of black holes}}}}\ (\bibinfo {year} {1985})\BibitemShut {NoStop}%
\bibitem [{\citenamefont {Jusufi}(2020{\natexlab{b}})}]{Jusufi:2020dhz}%
  \BibitemOpen
  \bibfield  {author} {\bibinfo {author} {\bibfnamefont {K.}~\bibnamefont
  {Jusufi}},\ }\href@noop {} {\  (\bibinfo {year} {2020}{\natexlab{b}})},\
  \Eprint {http://arxiv.org/abs/2004.04664} {arXiv:2004.04664 [gr-qc]}
  \BibitemShut {NoStop}%
\bibitem [{\citenamefont {Bisnovatyi-Kogan}\ and\ \citenamefont
  {Tsupko}(2017)}]{Bisnovatyi-Kogan:2017kii}%
  \BibitemOpen
  \bibfield  {author} {\bibinfo {author} {\bibfnamefont {G.~S.}\ \bibnamefont
  {Bisnovatyi-Kogan}}\ and\ \bibinfo {author} {\bibfnamefont {O.~Y.}\
  \bibnamefont {Tsupko}},\ }\href {\doibase 10.3390/universe3030057} {\bibfield
   {journal} {\bibinfo  {journal} {Universe}\ }\textbf {\bibinfo {volume}
  {3}},\ \bibinfo {pages} {57} (\bibinfo {year} {2017})},\ \Eprint
  {http://arxiv.org/abs/1905.06615} {arXiv:1905.06615 [gr-qc]} \BibitemShut
  {NoStop}%
\bibitem [{\citenamefont {Konoplya}\ and\ \citenamefont
  {Zhidenko}(2011)}]{Konoplya:2011qq}%
  \BibitemOpen
  \bibfield  {author} {\bibinfo {author} {\bibfnamefont {R.}~\bibnamefont
  {Konoplya}}\ and\ \bibinfo {author} {\bibfnamefont {A.}~\bibnamefont
  {Zhidenko}},\ }\href {\doibase 10.1103/RevModPhys.83.793} {\bibfield
  {journal} {\bibinfo  {journal} {Rev. Mod. Phys.}\ }\textbf {\bibinfo {volume}
  {83}},\ \bibinfo {pages} {793} (\bibinfo {year} {2011})},\ \Eprint
  {http://arxiv.org/abs/1102.4014} {arXiv:1102.4014 [gr-qc]} \BibitemShut
  {NoStop}%
\bibitem [{\citenamefont {Iyer}\ and\ \citenamefont
  {Will}(1987)}]{Iyer:1986np}%
  \BibitemOpen
  \bibfield  {author} {\bibinfo {author} {\bibfnamefont {S.}~\bibnamefont
  {Iyer}}\ and\ \bibinfo {author} {\bibfnamefont {C.~M.}\ \bibnamefont
  {Will}},\ }\href {\doibase 10.1103/PhysRevD.35.3621} {\bibfield  {journal}
  {\bibinfo  {journal} {Phys. Rev. D}\ }\textbf {\bibinfo {volume} {35}},\
  \bibinfo {pages} {3621} (\bibinfo {year} {1987})}\BibitemShut {NoStop}%
\bibitem [{\citenamefont {Kokkotas}\ and\ \citenamefont
  {Schutz}(1988)}]{Kokkotas:1988fm}%
  \BibitemOpen
  \bibfield  {author} {\bibinfo {author} {\bibfnamefont {K.}~\bibnamefont
  {Kokkotas}}\ and\ \bibinfo {author} {\bibfnamefont {B.~F.}\ \bibnamefont
  {Schutz}},\ }\href {\doibase 10.1103/PhysRevD.37.3378} {\bibfield  {journal}
  {\bibinfo  {journal} {Phys. Rev. D}\ }\textbf {\bibinfo {volume} {37}},\
  \bibinfo {pages} {3378} (\bibinfo {year} {1988})}\BibitemShut {NoStop}%
\bibitem [{\citenamefont {Seidel}\ and\ \citenamefont
  {Iyer}(1990)}]{Seidel:1989bp}%
  \BibitemOpen
  \bibfield  {author} {\bibinfo {author} {\bibfnamefont {E.}~\bibnamefont
  {Seidel}}\ and\ \bibinfo {author} {\bibfnamefont {S.}~\bibnamefont {Iyer}},\
  }\href {\doibase 10.1103/PhysRevD.41.374} {\bibfield  {journal} {\bibinfo
  {journal} {Phys. Rev. D}\ }\textbf {\bibinfo {volume} {41}},\ \bibinfo
  {pages} {374} (\bibinfo {year} {1990})}\BibitemShut {NoStop}%
\bibitem [{\citenamefont {Konoplya}(2003)}]{Konoplya_2003}%
  \BibitemOpen
  \bibfield  {author} {\bibinfo {author} {\bibfnamefont {R.}~\bibnamefont
  {Konoplya}},\ }\href {\doibase 10.1103/PhysRevD.68.024018} {\bibfield
  {journal} {\bibinfo  {journal} {Phys. Rev. D}\ }\textbf {\bibinfo {volume}
  {68}},\ \bibinfo {pages} {024018} (\bibinfo {year} {2003})},\ \Eprint
  {http://arxiv.org/abs/gr-qc/0303052} {arXiv:gr-qc/0303052} \BibitemShut
  {NoStop}%
\bibitem [{\citenamefont {Matyjasek}\ and\ \citenamefont
  {Opala}(2017)}]{Matyjasek:2017psv}%
  \BibitemOpen
  \bibfield  {author} {\bibinfo {author} {\bibfnamefont {J.}~\bibnamefont
  {Matyjasek}}\ and\ \bibinfo {author} {\bibfnamefont {M.}~\bibnamefont
  {Opala}},\ }\href {\doibase 10.1103/PhysRevD.96.024011} {\bibfield  {journal}
  {\bibinfo  {journal} {Phys. Rev. D}\ }\textbf {\bibinfo {volume} {96}},\
  \bibinfo {pages} {024011} (\bibinfo {year} {2017})},\ \Eprint
  {http://arxiv.org/abs/1704.00361} {arXiv:1704.00361 [gr-qc]} \BibitemShut
  {NoStop}%
\bibitem [{\citenamefont {Tangherlini}(1963)}]{Tangherlini:1963bw}%
  \BibitemOpen
  \bibfield  {author} {\bibinfo {author} {\bibfnamefont {F.}~\bibnamefont
  {Tangherlini}},\ }\href {\doibase 10.1007/BF02784569} {\bibfield  {journal}
  {\bibinfo  {journal} {Nuovo Cim.}\ }\textbf {\bibinfo {volume} {27}},\
  \bibinfo {pages} {636} (\bibinfo {year} {1963})}\BibitemShut {NoStop}%
\bibitem [{\citenamefont {Carter}(1968)}]{PhysRev.174.1559}%
  \BibitemOpen
  \bibfield  {author} {\bibinfo {author} {\bibfnamefont {B.}~\bibnamefont
  {Carter}},\ }\href {\doibase 10.1103/PhysRev.174.1559} {\bibfield  {journal}
  {\bibinfo  {journal} {Phys. Rev.}\ }\textbf {\bibinfo {volume} {174}},\
  \bibinfo {pages} {1559} (\bibinfo {year} {1968})}\BibitemShut {NoStop}%
\bibitem [{\citenamefont {Singh}\ and\ \citenamefont
  {Ghosh}(2018)}]{Singh_2018}%
  \BibitemOpen
  \bibfield  {author} {\bibinfo {author} {\bibfnamefont {B.~P.}\ \bibnamefont
  {Singh}}\ and\ \bibinfo {author} {\bibfnamefont {S.~G.}\ \bibnamefont
  {Ghosh}},\ }\href {\doibase 10.1016/j.aop.2018.05.010} {\bibfield  {journal}
  {\bibinfo  {journal} {Annals of Physics}\ }\textbf {\bibinfo {volume}
  {395}},\ \bibinfo {pages} {127–137} (\bibinfo {year} {2018})}\BibitemShut
  {NoStop}%
\bibitem [{\citenamefont {Zhidenko}(2006)}]{Zhidenko:2006rs}%
  \BibitemOpen
  \bibfield  {author} {\bibinfo {author} {\bibfnamefont {A.}~\bibnamefont
  {Zhidenko}},\ }\href {\doibase 10.1103/PhysRevD.74.064017} {\bibfield
  {journal} {\bibinfo  {journal} {Phys. Rev. D}\ }\textbf {\bibinfo {volume}
  {74}},\ \bibinfo {pages} {064017} (\bibinfo {year} {2006})},\ \Eprint
  {http://arxiv.org/abs/gr-qc/0607133} {arXiv:gr-qc/0607133} \BibitemShut
  {NoStop}%
\bibitem [{\citenamefont {Kiselev}(2003)}]{Kiselev:2002dx}%
  \BibitemOpen
  \bibfield  {author} {\bibinfo {author} {\bibfnamefont {V.}~\bibnamefont
  {Kiselev}},\ }\href {\doibase 10.1088/0264-9381/20/6/310} {\bibfield
  {journal} {\bibinfo  {journal} {Class. Quant. Grav.}\ }\textbf {\bibinfo
  {volume} {20}},\ \bibinfo {pages} {1187} (\bibinfo {year} {2003})},\ \Eprint
  {http://arxiv.org/abs/gr-qc/0210040} {arXiv:gr-qc/0210040} \BibitemShut
  {NoStop}%
\bibitem [{\citenamefont {Cuadros-Melgar}\ \emph {et~al.}(2020)\citenamefont
  {Cuadros-Melgar}, \citenamefont {Fontana},\ and\ \citenamefont
  {de~Oliveira}}]{Cuadros-Melgar:2020shz}%
  \BibitemOpen
  \bibfield  {author} {\bibinfo {author} {\bibfnamefont {B.}~\bibnamefont
  {Cuadros-Melgar}}, \bibinfo {author} {\bibfnamefont {R.}~\bibnamefont
  {Fontana}}, \ and\ \bibinfo {author} {\bibfnamefont {J.}~\bibnamefont
  {de~Oliveira}},\ }\href@noop {} {\  (\bibinfo {year} {2020})},\ \Eprint
  {http://arxiv.org/abs/2003.00564} {arXiv:2003.00564 [gr-qc]} \BibitemShut
  {NoStop}%
\end{thebibliography}%

\end{document}